\documentclass[sigconf]{acmart}

\usepackage{booktabs}
\usepackage{multirow}
\usepackage[most]{tcolorbox}
\usepackage{tabularx}
\usepackage[ruled,vlined]{algorithm2e}
\usepackage{xcolor}
\usepackage{threeparttable}
\usepackage{fancyhdr}
\usepackage{microtype}  % 改善文本排版
\usepackage{hyphenat}   % 改善断字

\newcommand{\lightcomment}[1]{\textcolor{gray}{#1}}
\newcolumntype{C}{>{\centering\arraybackslash}X}
\newcolumntype{L}{>{\raggedright\arraybackslash}X}

\newtcolorbox{promptbox}[1][]{
    colback=blue!5,
    colframe=blue!75!black,
    fonttitle=\bfseries,
    title=Prompt
}

\newtcolorbox{responsebox}[1][]{
    colback=green!5,
    colframe=green!75!black,
    fonttitle=\bfseries,
    title=AI
}

\SetKwSty{textbf}
\SetArgSty{textnormal}

\SetKwFor{For}{for}{:}{end}
\SetKwFor{While}{while}{:}{end}
\SetKwIF{If}{ElseIf}{Else}{if}{:}{elif:}{else:}{end}

\setcopyright{none}

\renewcommand\footnotetextcopyrightpermission[1]{}

\makeatletter
\fancypagestyle{myplain}{%
    \fancyhf{}
    \fancyhead[L]{\footnotesize Revisiting Vulnerability Patch Localization: An Empirical Study and LLM-Based Solution}
    \fancyhead[R]{\thepage}

}
\makeatother

\AtBeginDocument{%
  }

\begin{document}

\title{Revisiting Vulnerability Patch Localization: An Empirical Study and LLM-Based Solution}

% Xu Haoran, Zhang Yu, Zhi Chen, Junxiao Han, Xinkui Zhao, Jianwei Yin, Shuiguang Deng
\author{Haoran Xu}
\affiliation{%
  \institution{Zhejiang University}
  % \streetaddress{1 Th{\o}rv{\"a}ld Circle}
  \city{Ningbo}
  \country{China}}
\email{haoran.x@zju.edu.cn}

\author{Chen Zhi}
\authornote{Corresponding authors.}
\affiliation{ 
  \institution{Zhejiang University}
  % \streetaddress{1 Th{\o}rv{\"a}ld Circle}
  \city{Hangzhou}
  \country{China}}
\email{zjuzhichen@zju.edu.cn}

\author{Junxiao Han}
\affiliation{%
  \institution{Hangzhou City University}
  % \streetaddress{1 Th{\o}rv{\"a}ld Circle}
  \city{Hangzhou}
  \country{China}}
\email{hanjx@hzcu.edu.cn}

\author{Xinkui Zhao}
\affiliation{%
  \institution{Zhejiang University}
  % \streetaddress{1 Th{\o}rv{\"a}ld Circle}
  \city{Ningbo}
  \country{China}}
\email{zhaoxinkui@zju.edu.cn}

\author{Jianwei Yin}
\affiliation{%
  \institution{Zhejiang University}
  % \streetaddress{1 Th{\o}rv{\"a}ld Circle}
  \city{Hangzhou}
  \country{China}}
\email{zjuyjw@cs.zju.edu.cn}

\author{Shuiguang Deng}
\affiliation{%
  \institution{Zhejiang University}
  % \streetaddress{1 Th{\o}rv{\"a}ld Circle}
  \city{Hangzhou}
  \country{China}}
\email{dsg@zju.edu.cn}

\renewcommand{\shortauthors}{Xu et al.}

\begin{abstract}
Open-source software vulnerability patch detection is a critical component for maintaining software security and ensuring software supply chain integrity. Traditional manual detection methods face significant scalability challenges when processing large volumes of commit histories, while being prone to human errors and omissions. Existing automated approaches, including heuristic-based methods and pre-trained model solutions, suffer from limited accuracy, poor generalization capabilities, and inherent methodological constraints that hinder their practical deployment.
To address these fundamental challenges, this paper conducts a comprehensive empirical study of existing vulnerability patch detection methods, revealing four key insights that guide the design of effective solutions: the critical impact of search space reduction, the superiority of pre-trained semantic understanding over architectural complexity, the temporal limitations of web crawling approaches, and the advantages of knowledge-driven methods. Based on these insights, we propose a novel two-stage framework that combines version-driven candidate filtering with large language model-based multi-round dialogue voting to achieve accurate and efficient vulnerability patch identification. Extensive experiments on a dataset containing 750 real vulnerabilities demonstrate that our method outperforms current approaches.
\end{abstract}





\maketitle
\thispagestyle{empty}
\pagestyle{myplain}

\section{Introduction}
\label{sec:intro}
The expanding adoption of open-source software (OSS) \cite{tan2014bug} has brought increasing attention to OSS security vulnerabilities. As of December 2024, the National Vulnerability Database (NVD) \cite{nvd2025} has documented over 240,000 vulnerability entries, indicating a significant upward trend in OSS vulnerabilities. While these vulnerabilities are typically addressed through patches, many are not promptly reported to vulnerability databases or are silently fixed without public disclosure due to security concerns. This phenomenon allows other software to continue using vulnerable code, presenting opportunities for attackers. For instance, the Log4Shell vulnerability \cite{ibm2023log4j}, although discovered and reported to Apache on November 24, 2021, was not publicly disclosed until December 9, with the first patch released on December 10. This time lag between discovery, remediation, and patch deployment provides ample opportunity for attackers to develop and exploit vulnerabilities. Furthermore, even after patches are released, the remediation process for enterprises can take years, highlighting the critical importance of proactive discovery and rapid application of security patches. Traditionally, manually identifying vulnerability patches in software repositories is a time-consuming and error-prone task, proving exceptionally inefficient when searching for patches within large code repository. Consequently, research into automated techniques for detecting OSS vulnerability patch commits is of great significance, as it can not only enhance detection efficiency and ensure software product security but also supplement and improve vulnerability database information.

Existing methods for vulnerability patch detection primarily include traditional techniques based on fuzzing and symbolic execution, as well as approaches utilizing machine learning and deep learning. However, these methods exhibit limitations when dealing with incomplete code snippets or code changes within commits. Another category, reference-network-based methods, for example Tracer \cite{xu_tracking_2022}, have achieved some success in identifying vulnerability patches but overly rely on the accuracy and timeliness of reference links, discussions, and solutions, which can compromise the accuracy of detection results when processing newly disclosed vulnerability patches.

With the advent of large language models (LLMs), methods leveraging fine-tuned language models to pinpoint correct patches among candidate commits have also emerged. PatchFinder \cite{li_patchfinder_2024}, for example, first employs a semantic retriever to narrow down the search space by retrieving commits similar to the Common Vulnerabilities and Exposures (CVE) \cite{cve} description. Subsequently, a fine-tuned language model re-ranker is used to re-rank candidate commits based on relevance scores, prioritizing the most relevant patches.
PromVPat \cite{zhang_dual_2024}, on the other hand, utilizes a dual-prompt tuning channel to capture the semantic association between vulnerability descriptions and code commits, improving performance in few-shot scenarios. Specifically, its semantic matching module generates two new inputs using prompt templates and produces two relevance probabilities via a pre-trained language model (PLM) as semantic features. Concurrently, a handcrafted feature extraction module extracts 26 handcrafted features to capture explicit associations between vulnerability descriptions and code commits. Finally, the features from these two modules are integrated, the entire model is fine-tuned, and a classifier is used to predict relevance probabilities to locate security patches.

Among the aforementioned methods, reference-network-based approaches may overlook vulnerability patches not mentioned in the network, while most language-model-based methods rely on fine-tuned models. To avoid missing certain vulnerability patches and to save the time and cost associated with model fine-tuning, this paper proposes an automated method for detecting vulnerability patch commits based on version patch commit filtering and LLMs. Leveraging the powerful generalization capabilities and superior processing performance of LLMs in natural language processing (NLP), combined with chain-of-thought (CoT) \cite{wei2022chain} design and a multi-round voting system in prompt engineering, our method further optimizes the identification and verification of patch commits. Experimental results demonstrate that this method outperforms existing approaches in terms of accuracy, recall, and the number of successfully found patch commits, proving its effective applicability in practical software projects. Our specific contributions are as follows:
\begin{itemize}
    \item \textbf{Real-world Dataset Construction}: We collected vulnerability and corresponding patch information from real-world vulnerability management platforms to evaluate the practical performance of existing methods.
    \item \textbf{Empirical Study and Insights}: We conducted an empirical study of four vulnerability localization methods on real-world datasets, deriving four key insights about the limitations of existing methods and the actual data distribution characteristics in practical scenarios.
    \item \textbf{Two-Stage Detection Framework}: Based on our insights, we propose a two-stage framework that filters code commits based on version information and matches candidate commits through large language model multi-round dialogue, with comprehensive experimental validation demonstrating its effectiveness.
\end{itemize}

The remainder of this paper is organized as follows. Section \ref{sec:backgroud} presents an empirical study that replicates existing vulnerability patch detection methods on real-world datasets, revealing key limitations and deriving actionable insights. Section \ref{sec:approach} introduces our two-stage framework designed to address the identified challenges. Section \ref{sec:eval} compares the optimal method from the empirical study with our proposed method.

\section{Empirical Study}
\label{sec:backgroud}
Existing vulnerability patch localization methods offer various theoretical solutions, but they face numerous challenges in practical deployment and use. To gain a deeper understanding of their limitations and provide an empirical basis for improvement, this section systematically evaluates the practical effectiveness of current mainstream methods through large-scale data analysis and experimental validation. We aim to answer the following research questions:
 
\begin{itemize}
    \item \textbf{RQ1:} Why are current vulnerability patch localization methods ineffective in practice?
    \item \textbf{RQ2:} What insights can we gain from a systematic observation of the vulnerabilities?
\end{itemize}

\subsection{Experiment Setup}
\subsubsection{Baseline}
We selected a total of four methods: two recent vulnerability localization approaches \cite{li_patchfinder_2024, zhang_dual_2024}, one traditional feature-based method \cite{wang_vcmatch_2022}, and one traditional web search-based method \cite{xu_tracking_2022}. Among these, PatchFinder \cite{li_patchfinder_2024} and PromVPat \cite{zhang_dual_2024} are language model fine-tuning-based methods, while VCMatch \cite{wang_vcmatch_2022} is a feature learning-based approach. We reproduced these methods using their open-source code and data, employing the default hyperparameter settings provided in the open-source implementations. 
For the web search method Tracer \cite{xu_tracking_2022}, since the original homepage \cite{tracer_page_2022} has discontinued code release, we obtained a forked version of the open-source code from another repository \cite{tracer_code_2021}. We refactored the code to ensure proper functionality \cite{VulnerCollector_2023}.
During experimentation, we evaluated metrics by comparing the patch links crawled by Tracer from the internet against our dataset. It should be noted that Tracer may ultimately retrieve URLs that are equivalent to but different from the annotated patch URLs in the dataset (different link path formats that ultimately point to the same webpage). Therefore, we evaluated patch consistency based on commit hashes rather than URL strings.

\subsubsection{Dataset}
\label{sec:dataset}
We initially referenced baseline datasets, partially utilizing the deep vulnerability dataset constructed by Xu et al. \cite{xu_tracking_2022} as the foundation for our research dataset. This 
vulnerabilities from 2010 to 2020, along with manually validated patch commits, primarily covering multiple programming languages including Java, Python, C++, and Go. Additionally, we crawled open-source vulnerabilities from VERA \cite{vera} and SNYK \cite{snyk_vuln_db} spanning 2020 to 2024, subsequently extracting open-source software vulnerabilities containing identical vulnerability patch commits and manually verifying the correctness of these vulnerability patch commits. Ultimately, we obtained a subset of open-source software vulnerabilities from 2020 to 2024 along with their corresponding patch commit data. We then cloned the GitHub repositories affected by these vulnerabilities and, based on the version tag ranges where patches were located, extracted all commits within these version ranges to verify whether baselines could identify the correct patch commits within this commit range.

\begin{figure}[t]
    \centering
    \includegraphics[width=0.9\linewidth]{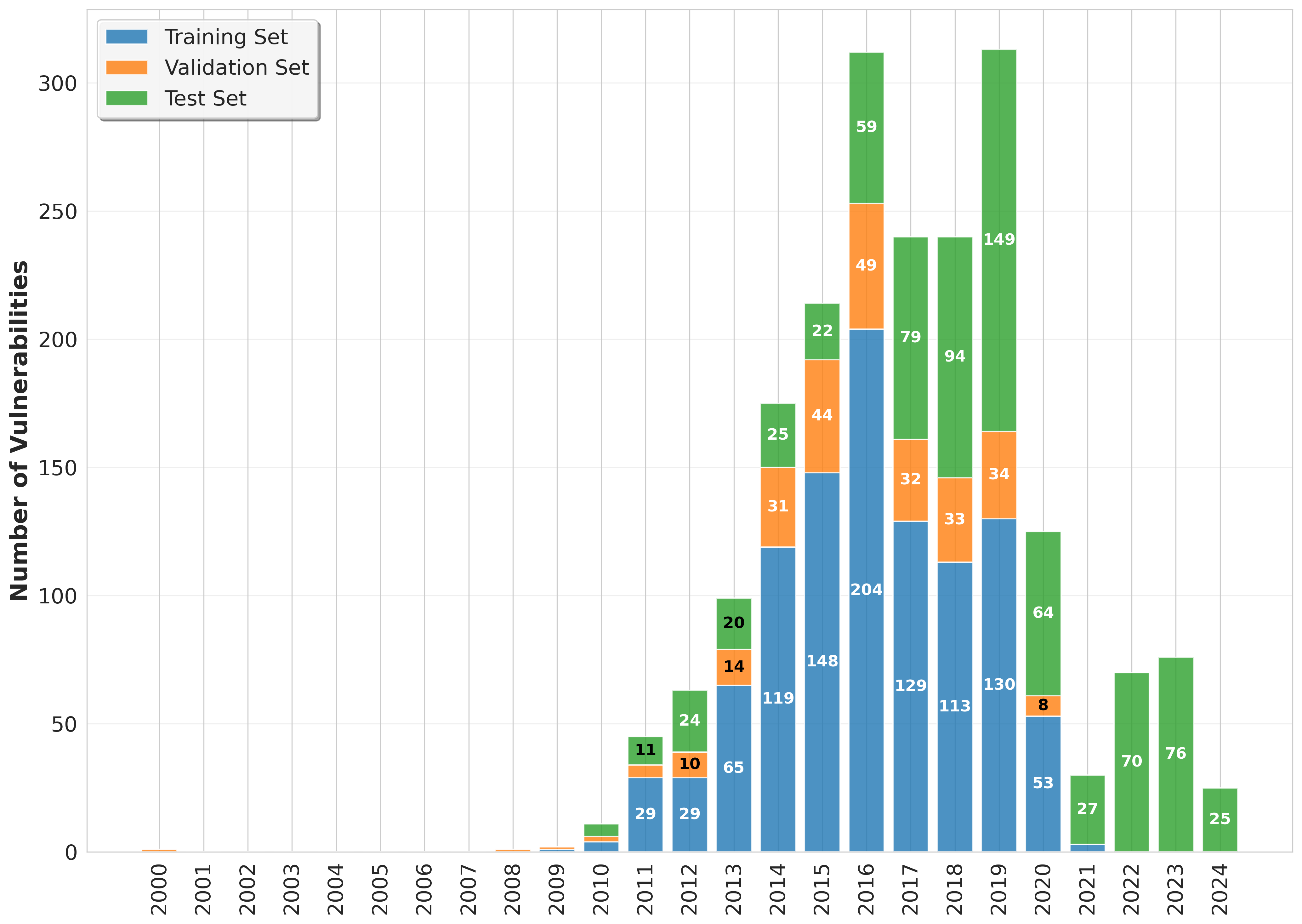}
    \caption{Vulnerability Distribution by Year}
    \label{fig:cve_year_distribution}
    \Description{Vulnerability Distribution by Year}
\end{figure}

Since Li et al. \cite{li_patchfinder_2024} did not fully release their dataset, we selected Zhang et al.'s dataset \cite{zhang_dual_2024} to train the baselines, whose data originated from VCMatch \cite{wang_vcmatch_2022} and SAP \cite{ponta2019manually}. We compared Zhang et al.'s dataset with our collected dataset, removed shared CVE vulnerabilities, and then split their dataset into training and validation sets at an 8:2 ratio for training PatchFinder, PromVPat, and VCMatch, while using our dataset as the test set. Finally, we filtered out entries with missing data and ensured that: (1) the training, validation, and test sets were mutually independent with no shared CVE vulnerabilities; (2) each CVE vulnerability contained at least one true positive patch; and (3) no training attributes in each data entry had missing values. Table \ref{tab:dataset_stat} and Figure \ref{fig:cve_year_distribution} provide a comprehensive statistical overview of our dataset.

\begin{table}[t]
\caption{The Statistics of Datasets}
\label{tab:dataset_stat}
\begin{tabular}{@{}l|llll@{}}
\toprule
Dataset         & Train  & Val   & Test   & Total   \\ \midrule
Vulnerability   & 1,027  & 265   & 750    & 2,042  \\
Commits         & 29,484 & 7670  & 61,780 & 98,934  \\
Average Commits & 28.71  & 28.94 & 82.37  & 48.45   \\
Database Size   & 30,310 & 7,876 & 73,841 & 112,027 \\ \bottomrule
\end{tabular}
\end{table}

\subsubsection{Metric}
Since these methods primarily focus on identifying relevant vulnerability patches, we employed recommendation domain-related metrics to reflect method performance, including average precision, recall, and F1-score for CVE vulnerability detection results. Specifically, precision refers to the proportion of correctly detected patch commits among all detected commits for each vulnerability; recall represents the proportion of correctly detected patch commits among all correct patches for each vulnerability in the dataset; and F1-score is the harmonic mean of precision and recall.

\begin{table*}[t]
\centering
\small
\caption{Performance of the replicated baseline method on a real-world dataset.}
\label{tab:baseline_result}
\begin{tabularx}{\textwidth}{lCCCCCCCCC}
\hline
\multirow{2}{*}{\textbf{Method}} & \multicolumn{3}{c}{\textbf{Top 1}}                  & \multicolumn{3}{c}{\textbf{Top 3}} & \multicolumn{3}{c}{\textbf{Top 5}} \\ \cline{2-10} 
                                 & Precision       & Recall          & F1              & Precision    & Recall   & F1       & Precision    & Recall   & F1       \\ \hline
PatchFinder                      & 0.2284          & 0.2095          & 0.2154          & 0.1547       & 0.4224   & 0.2230   & 0.1168       & 0.5307   & 0.1890   \\
PromVPat-mess                    & 0.0469          & 0.0391          & 0.0416          & 0.0749       & 0.2006   & 0.1072   & 0.0654       & 0.2951   & 0.1057   \\
PromVPat-code                    & 0.1235          & 0.1154          & 0.1179          & 0.1041       & 0.2856   & 0.1504   & 0.0857       & 0.3887   & 0.1386   \\
PromVPat                         & 0.0296          & 0.0249          & 0.0263          & 0.0609       & 0.1685   & 0.0883   & 0.0548       & 0.2506   & 0.0889   \\
VCMatch                          & 0.0360          & 0.0289          & 0.0306          & 0.0716       & 0.1592   & 0.0938   & 0.0744       & 0.2565   & 0.1094   \\ \hline
Tracer                           & \textbf{0.5303} & \textbf{0.5734} & \textbf{0.5235} & -            & -        & -        & -            & -        & -        \\ \hline
\end{tabularx}
\end{table*}

\subsection{Effectiveness of Existing Methods}
We evaluated these baselines on our collected dataset to locate correct vulnerability patches within a version range. Tracer is an exception, as it identifies corresponding vulnerability patches through web search based on CVE numbers. We assessed the performance of multiple methods across different Top-K settings, including PatchFinder, PromVPat and its variants, VCMatch, and Tracer. We compared the precision, recall, and F1 scores of these methods at Top-1, Top-3, and Top-5 recommendations. The experimental results are shown in Table \ref{tab:baseline_result}. The implementation of each method in the table is briefly described as follows:

\textbf{PatchFinder} is an end-to-end fine-tuning architecture based on CodeReviewer \cite{li_code_review}, which utilizes CodeReviewer to encode CVE descriptions and commits, calculating similarity scores to rank commits. 

\textbf{PromVPat} employs CodeT5 \cite{wang2021codet5} encoding combined with feature encoding to rank commits, separately encoding and classifying commit messages and code, then incorporating handcrafted features, and finally fusing the three encoding components through an attention layer to score commits. 

\textbf{PromVPat-mess} and \textbf{PromVPat-code} represent the scenarios where PromVPat uses only the commit message or code encoder for scoring, respectively. 

\textbf{VCMatch} utilizes handcrafted feature vectors and textual feature vectors of vulnerabilities and commits, obtaining rankings through various machine learning methods, and finally integrates all method results using a voting approach. 

\textbf{Tracer} constructs a vulnerability-related web graph through deep crawling of vulnerability website pages, calculates the relevance of related commit pages, and identifies corresponding vulnerability patches. Since Tracer directly obtains vulnerability patches without Top-K differentiation, it yields final precision, recall, and F1 scores directly.

\subsection{Scaling Challenges of Baseline Methods}
As software projects continue to evolve, the number of commits in code repositories exhibits rapid growth trends, causing machine learning-based vulnerability patch localization methods to demonstrate significant performance limitations in practical applications. Through systematic evaluation of existing baseline methods, we found that the localization accuracy of all methods falls far below the threshold required for real-world applications, reflecting underlying methodological issues.

We conducted a unified evaluation of five feature-based training methods using identical datasets and evaluation metrics to ensure result comparability. Experimental results show that even the best-performing method achieves only 22.84\% Top-1 accuracy, which is substantially below the threshold required for practical applications. This widespread performance deficiency prompts us to investigate the fundamental causes. We hypothesize that the \textbf{excessively large search space of candidate commits is the primary factor contributing to the poor performance of existing methods}.

\subsubsection{Analysis of Candidate Commit Scale within Version Ranges}
As software projects continue to evolve, the number of commits in code repositories exhibits rapid growth trends, causing machine learning-based vulnerability patch localization methods to demonstrate significant performance limitations in practical applications. Through systematic evaluation of existing baseline methods, we found that the localization accuracy of all methods falls far below the threshold required for real-world applications, reflecting underlying methodological issues.

We conducted a unified evaluation of five feature-based training methods using identical datasets and evaluation metrics to ensure result comparability. Experimental results show that even the best-performing method achieves only 23\% Top-1 accuracy, which is substantially below the threshold required for practical applications. This widespread performance deficiency prompts us to investigate the fundamental causes. We hypothesize that the excessively large search space of candidate commits is the primary factor contributing to the poor performance of existing methods.

To validate that the oversized search space is the primary cause of poor performance in training-based methods, we conducted a systematic analysis of the candidate commit scale corresponding to 10,407 CVE instances collected from Vera \cite{vera} and Snyk \cite{snyk_vuln_db} databases. We consider all commits within the patch version range as candidate commits. These data span multiple typical open-source projects, providing a comprehensive empirical foundation for understanding search complexity in real-world application scenarios.

% \begin{figure*}[t]
%     \centering
%     \includegraphics[width=0.9\linewidth]{pic/cloc_analyzer2.1_2.png}
%     \caption{Distribution of Commits with Different Patch Counts}
%     \label{fig:cloc_analyzer_2}
% \end{figure*}

\subsubsection{Prevalence of Large-Scale Search Spaces}
\begin{figure*}[t]
    \centering
    \includegraphics[width=0.9\linewidth]{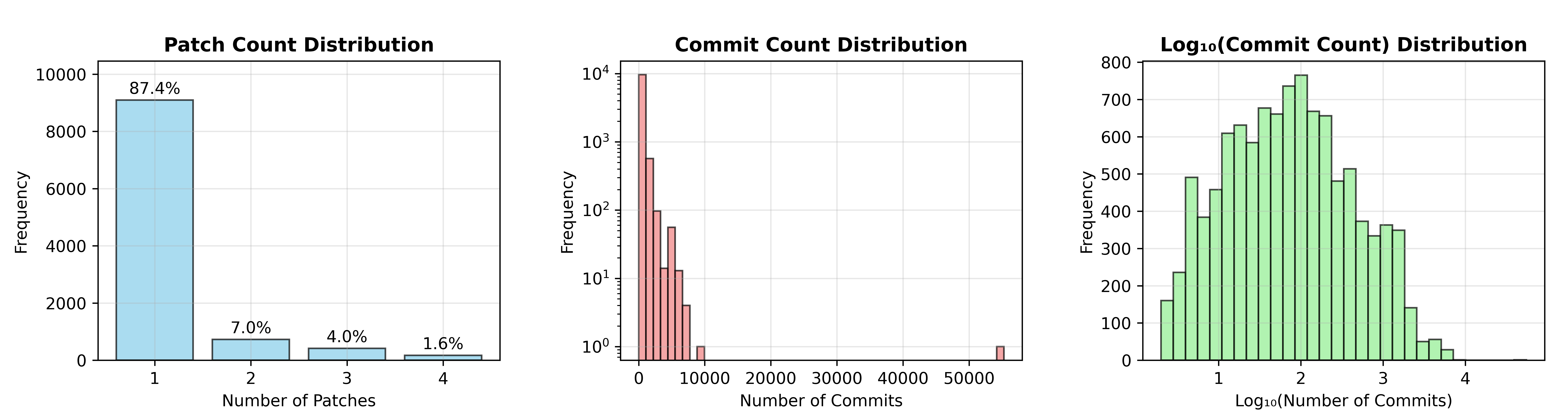}
    \caption{Patch and Candidate Commit Distribution}
    \label{fig:cloc_analyzer_1}
    \Description{Patch and Candidate Commit Distribution}
\end{figure*}

The specific distribution is shown in Figure \ref{fig:cloc_analyzer_1}. 
% To compare the distribution of different patch quantities, we also generated violin plots for commits with different patch counts, as illustrated in Figure \ref{fig:cloc_analyzer_2}.
Our analysis reveals that the number of candidate commits corresponding to each CVE exhibits extremely uneven distribution characteristics: each vulnerability averages 302 candidate commits with a median of 70, but the standard deviation reaches 836. The distribution of candidate commit quantities shows severe right-skewness (skewness coefficient = 29.467), leading to the following conclusions:

\begin{itemize}
    \item \textbf{Significant long-tail effect}: While most CVEs have relatively few candidate commits, a considerable number of extreme cases exist.
    \item \textbf{Mean distortion}: Due to extreme values, the arithmetic mean (302) far exceeds the median (70), failing to accurately reflect typical scenarios.
    \item \textbf{Polarized search difficulty}: Approximately 25\% of CVEs face relatively simple search tasks (<18 candidates), while another 25\% encounter extremely complex search scenarios (>252 candidates).
\end{itemize}
These data demonstrate the prevalence of large-scale search spaces. Large-scale search spaces are not marginal phenomena but common challenges in vulnerability patch localization tasks.

\begin{table}[t]
\centering
\caption{Top 5 Cases with Largest Candidate Commit Counts}
\label{tab:extreme_cases}
\begin{tabular}{lrl}
\toprule
\textbf{Project} & \textbf{Commits} % & \textbf{Patches} 
& \textbf{Project Type} \\
\midrule
pfsense/FreeBSD-ports & 55,276 % & 1 
& System Software \\
magento/magento2 & 9,251 % & 1 
& Web Application \\
istio/envoy & 7,042 % & 1 
& Network Proxy \\
microweber/microweber & 6,749 % & 1 
& CMS\textsuperscript{*} \\
zephyrproject-rtos/zephyr & 6,360 % & 1 
& Embedded OS \\
\bottomrule
\end{tabular}
\begin{tablenotes}
\footnotesize
\item[*] CMS: Content Management System
\end{tablenotes}
\end{table}

% \begin{table}[]
% \centering
% \caption{CVE Analysis of Popular Open Source Projects}
% \label{tab:cve_analysis}
% \begin{tabular}{lrrr}
% \toprule
% \textbf{Project} & \textbf{Total Commits} & \textbf{CVEs} & \textbf{Avg Commits/CVE} \\
% \midrule
% torvalds/linux & 373,802 & 204 & 1,832 \\
% moodle/moodle & 365,925 & 315 & 1,162 \\
% tensorflow/tensorflow & 345,581 & 199 & 1,737 \\
% gpac/gpac & 116,713 & 120 & 973 \\
% microweber/microweber & 91,728 & 67 & 1,369 \\
% zephyrproject-rtos/zephyr & 91,393 & 22 & 4,154 \\
% Dolibarr/dolibarr & 67,917 & 60 & 1,132 \\
% pfsense/FreeBSD-ports & 55,276 & 1 & 55,276 \\
% CGAL/cgal & 49,680 & 45 & 1,104 \\
% FFmpeg/FFmpeg & 46,719 & 47 & 994 \\
% \bottomrule
% \end{tabular}
% \end{table}

\subsubsection{Search Challenges in Ultra-Large-Scale Projects}
According to our statistics, we identified several extreme search space examples, as shown in Table \ref{tab:extreme_cases}. This scale distribution directly impacts the performance of training-based methods. Taking the pfsense/FreeBSD-ports project as an example, a single CVE corresponds to 55,276 candidate commits, meaning the model must accurately identify the unique patch commit among tens of thousands of samples. Similarly, mainstream projects like magento/magento2 and tensorflow/tensorflow frequently involve thousands of candidate commits per CVE. 
%Based on the statistics in Table \ref{tab:cve_analysis},
We observe that large-scale projects typically exhibit higher search complexity, and active, large-scale projects also contain more vulnerabilities and patches. Large system software projects like the Linux kernel average 1,832 candidate commits per CVE, while machine learning frameworks like TensorFlow average 1,737. In contrast, some specialized software projects, despite having fewer CVEs, present equally formidable search complexity for individual vulnerabilities. These inter-project differences reflect the diversity of software ecosystems while demonstrating the universality of oversized search spaces. 

Our analysis reveals significant differences in search complexity across different project types. For instance, system software projects (such as Linux and FreeBSD) typically have average candidate commit counts of 1,500+, exhibiting high mean and high variance distributions. The challenge lies in their long-term maintenance and extensive commit histories. Web application projects (such as Moodle and Magento) generally have average candidate commit counts ranging from 800-1,200, with challenges stemming from rapid iteration and large version spans. Specialized software projects (such as embedded systems) have smaller average candidate commit counts but may suffer from irregular version management practices. Whether continuously maintained large-scale projects or rapidly iterating application software, all face the challenge of enormous candidate commit quantities.

\begin{center}
\begin{tcolorbox}[colback=gray!20,%gray background
                  colframe=black,% black frame colour
                  arc=1mm, auto outer arc,
                  boxrule=0.5pt,
                 ]

\textbf{Insight 1: }
\textbf{The oversized search space represents the bottleneck constraining the performance of feature-based training methods, rather than being simply a model design issue.} When the number of candidate commits reaches hundreds or even thousands, any machine learning model struggles to accurately locate patches within extremely imbalanced samples. Therefore, effective solutions should begin with reducing the search space.
\end{tcolorbox}
\end{center}

\subsection{Timeliness Issues in Crawler-Based Methods}
From the experimental results, we observe that although Tracer demonstrates significantly better performance compared to other methods, achieving substantial improvements in patch accuracy, recall, and F1-score while reducing the manual workload for vulnerability patch localization, it also has its limitations. Information retrieval-based methods rely on obtaining patch clues from developer discussions, public datasets, issue tracking systems, and other channels. However, the availability and timeliness of such information have inherent limitations. For instance, when reference materials are of poor quality or published with delays, Tracer's detection effectiveness may be impacted. Through analysis of Tracer's historical data, we found that it exhibits certain delay issues.

\begin{figure}[t]
    \centering
    \includegraphics[width=0.9\linewidth]{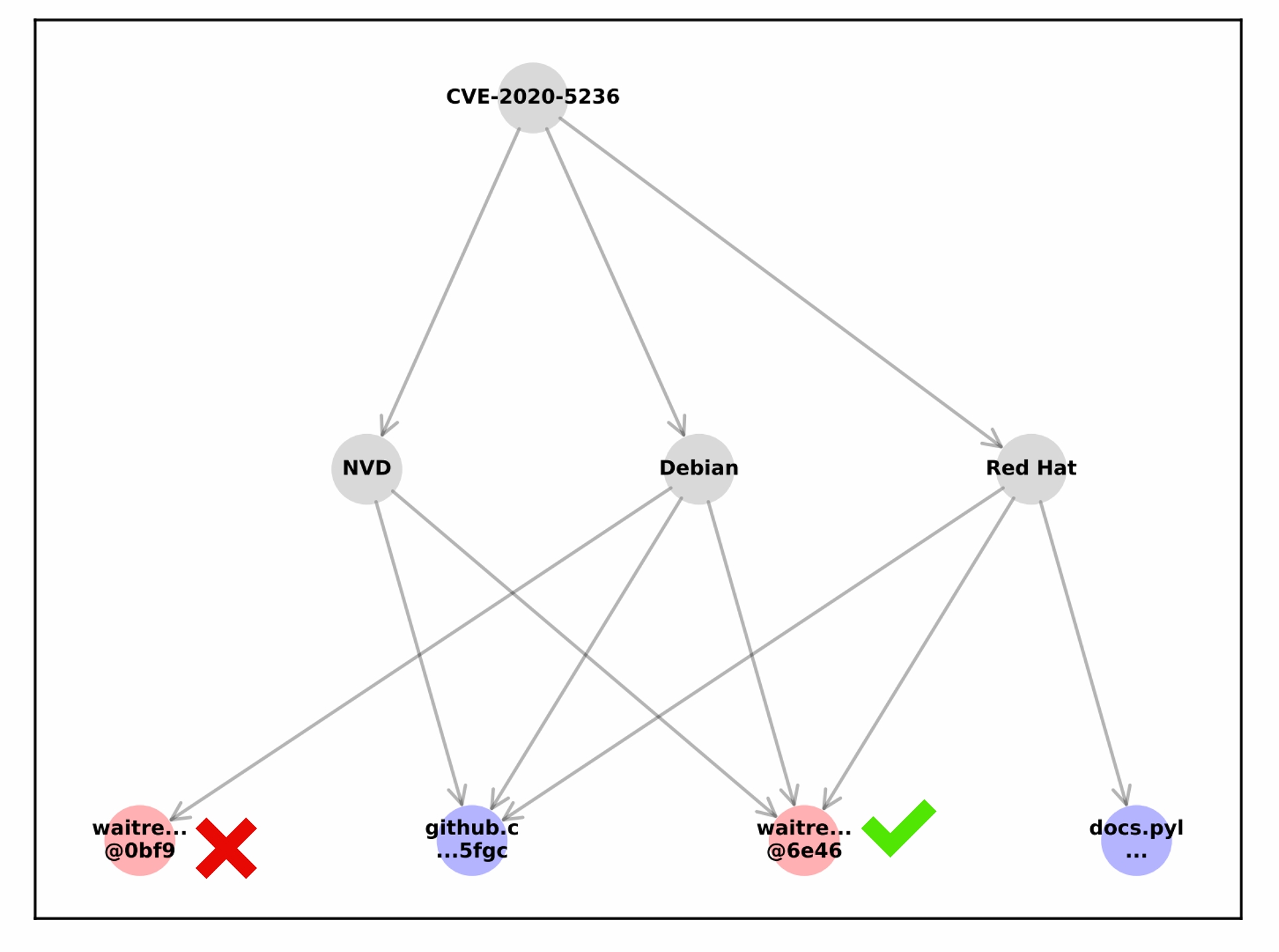}
    \caption{Reference Network for CVE-2020-5236}
    \label{fig:CVE-2020-5236}
    \Description{Reference Network for CVE-2020-5236}
\end{figure}

For example, when we reproduced Tracer using 1,295 open-source software vulnerabilities from our comprehensive dataset, we discovered that 59 vulnerability patches had reference paths with delays exceeding 100 days, while 139 vulnerabilities had patch reference paths with delays ranging from 10 to 100 days. Taking vulnerability CVE-2020-5236 as an example, base on Tracer's constructed reference network, 
shown in Figure \ref{fig:CVE-2020-5236}, 
where Tracer selected commit 6e46f9e as the final patch commit. However, Tracer could only rely on the vulnerability advisory published by RedHat (released on May 5, 2020), but this advisory's publication date was later than the actual time of patch commit 6e46f9e (February 3, 2020), with a time difference of 92 days. The existence of such time delays indicates that in certain cases, Tracer may not be able to timely utilize the most accurate reference materials, resulting in limitations to the effectiveness of its tracking results.

\begin{center}
\begin{tcolorbox}[colback=gray!20,%gray background
                  colframe=black,% black frame colour
                  arc=1mm, auto outer arc,
                  boxrule=0.5pt,
                 ]

\textbf{Insight 2: }
\textbf{While crawler-based vulnerability patch localization methods demonstrate high accuracy, they suffer from inherent timeliness limitations}, the dependency on delays in third-party information publication makes them unable to meet the demands of real-time vulnerability response. This highlights the importance of developing autonomous methods capable of independently identifying patches without relying solely on external information sources.
\end{tcolorbox}
\end{center}

\subsection{Feature Representation Effectiveness Issues}
The PromVPat series results reveal critical insights about feature fusion strategies. Notably, code features prove more effective than message features (PromVPat-code F1: 0.1179 vs. PromVPat-mess F1: 0.0416), reflecting the essential nature of vulnerability patch identification: patches fundamentally involve code modifications, while commit messages are often concise or ambiguous. More significantly, the complete PromVPat shows degraded performance compared to PromVPat-code alone (F1 drops from 0.1179 to 0.0263), revealing a "noise amplification" effect in feature fusion. This degradation likely stems from distributional differences among feature types causing fusion bias, or attention mechanisms failing to learn appropriate feature importance weightings.

Among feature-based methods, PatchFinder demonstrates superior performance despite its conceptual simplicity. While VCMatch employs complex multi-method voting and PromVPat uses sophisticated attention-based fusion, both show limited effectiveness compared to PatchFinder's straightforward approach. This suggests that VCMatch's ensemble complexity may amplify noise in small-sample, high-noise vulnerability data rather than improving robustness. PatchFinder's success likely benefits from CodeReviewer's large-scale code pre-training, providing robust code comprehension capabilities that outweigh architectural sophistication.

\begin{center}
\begin{tcolorbox}[colback=gray!20,%gray background
                  colframe=black,% black frame colour
                  arc=1mm, auto outer arc,
                  boxrule=0.5pt,
                 ]

\textbf{Insight 3: }
\textbf{In vulnerability patch localization, pre-trained semantic understanding outperforms architectural complexity.} Simple feature-based methods may outperform complex feature fusion and ensemble approaches, indicating that domain-specific pre-trained semantic understanding is more valuable than sophisticated ensemble strategies in noisy real-world datasets.
\end{tcolorbox}
\end{center}

\subsection{Fundamental Differences in Information Acquisition Dimensions}
From the experimental results, we observe that \textbf{the Tracer method demonstrates superior performance}, achieving 52.81\% precision, 57.09\% recall, and 52.13\% F1-score on the Top-1 metric, significantly outperforming all other machine learning-based methods. This indicates that the approach of constructing vulnerability-related graphs through web crawling possesses distinct advantages in vulnerability patch localization tasks.

We believe the most fundamental difference between Tracer and other methods lies in \textbf{different dimensions of information acquisition}. Machine learning methods primarily rely on textual feature matching between CVE descriptions and commit content, which has inherent limitations: CVE descriptions are often abstract and general, while actual vulnerability patches may involve specific implementation details, creating a "semantic gap" between them. In contrast, Tracer constructs association graphs through deep crawling of vulnerability websites, \textbf{leveraging rich contextual information from the web}, including vulnerability discussions, related links, developer communications, and more. This information often contains the complete chain from vulnerability discovery to patch remediation, providing more direct and accurate association cues.

This represents a \textbf{fundamental methodological difference}. Traditional machine learning methods are essentially \textbf{data-driven}, attempting to learn vulnerability-patch mapping relationships from limited training samples. However, vulnerability patch localization is a task that heavily depends on domain knowledge and contextual information, where pure statistical learning may struggle to capture complex associations.

Tracer's success likely stems from its \textbf{knowledge-driven} nature: by leveraging structured knowledge and manually curated information from the internet, it bypasses the difficulty of learning complex mappings from raw data. This approach more closely resembles how human experts solve such problems—relying on rich background knowledge and comprehensive judgment from multiple information sources.

This comparative result suggests that in certain highly specialized tasks, fully leveraging external knowledge sources may be more effective than pure machine learning modeling.

It is worth noting that large-scale pre-trained language models can also be viewed as a knowledge-driven approach. These models have internalized rich domain knowledge and contextual understanding capabilities through pre-training on massive text and code data. Compared to traditional machine learning methods, large models can understand the complex relationships between code semantics, vulnerability patterns, and remediation strategies without learning these mappings from scratch.

From this perspective, PatchFinder's relative success based on the CodeReviewer pre-trained model is not coincidental—it actually leverages the knowledge learned by the model on large-scale code corpora. This approach sits between purely data-driven and explicitly knowledge-driven methods, obtaining rich background knowledge similar to Tracer through pre-training, but stored in parameterized form within the model.

Therefore, utilizing large language models for vulnerability patch localization may be a more promising direction than traditional feature engineering. Large models not only possess powerful semantic understanding capabilities but can also process multimodal information (code, documentation, discussions, etc.). More importantly, they can perform few-shot learning and reasoning—characteristics that are well-suited for vulnerability patch localization tasks requiring deep understanding and complex reasoning.

\begin{center}
\begin{tcolorbox}[colback=gray!20,%gray background
                  colframe=black,% black frame colour
                  arc=1mm, auto outer arc,
                  boxrule=0.5pt,
                 ]

\textbf{Insight 4: }
\textbf{The superiority of knowledge-driven approaches over traditional feature-based methods} suggests that large language models, which internalize vast domain knowledge through pre-training, may represent the most promising direction for vulnerability patch localization—combining the semantic understanding of pre-trained models with the reasoning capabilities needed for complex code-vulnerability mappings.
\end{tcolorbox}
\end{center}

\subsection{Summary}
Based on the four core insights from the above analysis, we design a two-stage automated detection framework for open-source software vulnerability patches in Section \ref{sec:approach}, which specifically addresses the key limitations of existing methods.

To tackle the problem of oversized search spaces (\textbf{Insight 1}), our framework first reduces candidate commits through version information mining and multi-branch cross-filtering strategies, fundamentally resolving the performance bottleneck caused by excessive search spaces. This version-driven filtering mechanism avoids blind searching among massive irrelevant commits.

To overcome the timeliness limitations of crawler-based methods (\textbf{Insight 2}), our framework relies entirely on information from the code repository itself, requiring no dependency on third-party vulnerability advisories or external discussions, thus achieving autonomous patch identification. This design ensures real-time performance and independence, unaffected by publication delays of external information sources.

Based on the findings that pre-trained semantic understanding outperforms complex architectures and the advantages of knowledge-driven methods (\textbf{Insights 3 and 4}), we abandon complex feature fusion and ensemble strategies, directly leveraging the intrinsic semantic understanding capabilities of large language models for patch identification. Through carefully designed prompt engineering, we enable the model to fully utilize its internalized domain knowledge for complex code-vulnerability mapping reasoning. This approach combines the semantic understanding and logical reasoning capabilities of pre-trained models, achieving knowledge-driven intelligent patch localization.

\section{Approach}
\label{sec:approach}
\begin{figure*}[t]
    \centering
    \includegraphics[width=0.94\linewidth]{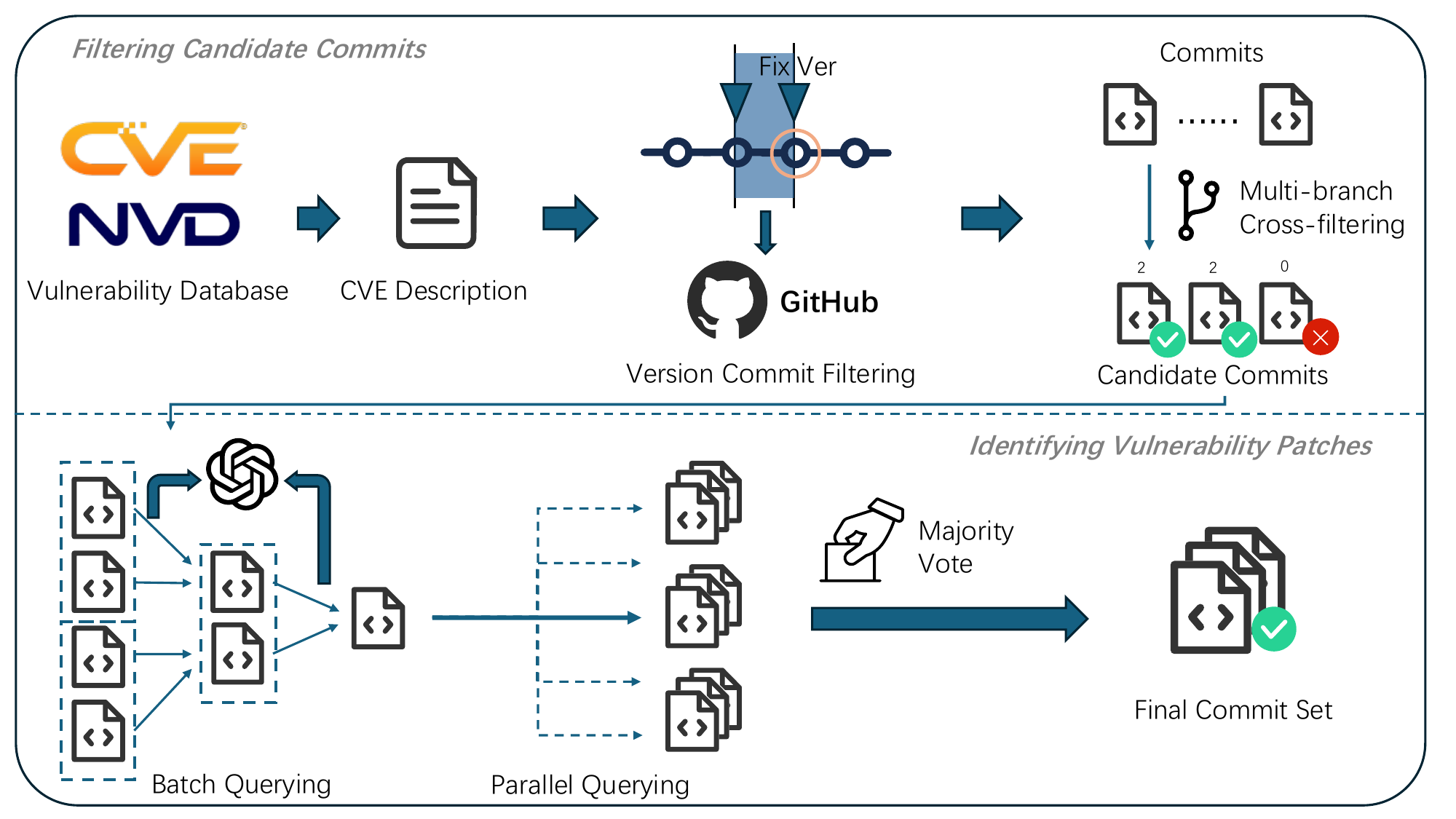}
    \caption{Two-stage open-source software vulnerability patch detection: Initially, affected versions are mined from vulnerability information to obtain candidate code commits, followed by multi-round conversational voting to select the definitive patch set.}
    \label{fig:framework}
    \Description{Two-stage open-source software vulnerability patch detection: Initially, affected versions are mined from vulnerability information to obtain candidate code commits, followed by multi-round conversational voting to select the definitive patch set.}
\end{figure*}

This study proposes a two-phase automated detection framework for open-source software vulnerability patches, as illustrated in Figure \ref{fig:framework}. The framework combines version commit filtering with multi-round dialogue voting to guide LLMs in accurately identifying the correct patch commits. This approach addresses key challenges in vulnerability patch detection, including incomplete patch information in vulnerability databases and the need to accurately identify corresponding patches within an extensive commit history. It helps system maintenance personnel become aware of vulnerability patch updates earlier, thereby preventing attacks on the system. The framework primarily consists of two phases: (1) Filtering Candidate Commits Phase: Information extraction and repository analysis are conducted to extract the software name and fixed version number associated with the CVE vulnerability, and candidate code commits are obtained through multi-branch cross-filtering; (2) Identifying Vulnerability Patches Phase: Key information from the code is extracted and prompts are constructed. A LLM is used to identify patches, and the final patch commit is determined through multi-round dialogue voting.

\subsection{Filtering Candidate Commits}
\label{sec:candidate_commits}
In this phase, the fixed version number is first identified based on the information from the vulnerability database. Then, the range of code commits is narrowed down. For repositories maintained with multiple branches, the Multi-branch Cross-filtering method is used to further filter potential patch commits.

\subsubsection{Version Acquisition}
The paper extracts relevant information for each vulnerability by mining multiple vulnerability databases such as CVE \cite{cve}, NVD \cite{nvd2025}, VERA \cite{vera}, and Snyk \cite{snyk_vuln_db}. The extracted information includes CVE ID, vulnerability description, CVSS score, CPE, etc. According to the characteristics of each dataset, the information that needs attention for each data item is shown in Table \ref{tab:vul_info}. Then, using NLP techniques, key information such as the software name indicated in the CPE, the version number of the vulnerability fix, and the programming language used in the code repository is extracted from the vulnerability descriptions. Next, based on the repository information of the GitHub project where the vulnerability resides, the commits and version tag information of the repository are obtained. Finally, the combined information yields the code repository and the version number of the vulnerability fix.

\begin{table}[t]
\caption{Vulnerability Information Table}
\label{tab:vul_info}
\begin{tabular}{@{}lcr@{}}
\toprule
\textbf{Database} & \textbf{Key Info} & \textbf{Description}          \\ \midrule
\multirow{2}{*}{CVE}            & CVD ID                & The unique identifier         \\
                                & Description          & A textual description         \\ \midrule
\multirow{3}{*}{NVD}            & CVSS                 & Vulnerability Severity Rating \\
                                & References           & Reference links               \\
                                & CPE                  & Vulnerability scope           \\ \midrule
\multirow{2}{*}{VERA}           & Patch                & Manually collected patches    \\
                                & UpdateToVersion      & Bug fix version number        \\ \midrule
\multirow{2}{*}{Snyk}           & References           & Reference links               \\
                                & PackageName          & Affected repositories         \\ \bottomrule
\end{tabular}
\end{table}

If the structured description of the vulnerability does not provide the associated repository and the fixed version, we design prompts as shown in Figure \ref{fig:prompt_ver} to query a LLM. This is used to determine the software name and the fixed version number based on the textual description of the CVE. However, our experiments indicate that the LLM may misidentify the software, for example, recognizing a single software as two different ones or mistaking function names in the description as software. Additionally, if version ranges are used in the descriptions, the LLM may not correctly return the version numbers within that range. Lastly, there are also some vulnerability descriptions that do not specify the affected versions.

\begin{figure}[t]

\centering
\begin{promptbox}
Your task is to find the affected repositories and the fixed version numbers based on the description of a vulnerability in the CVE information. There may be multiple repositories and version numbers. You need to list them in the format of (repository, version number).
Please carefully read the following CVE description:

\textless cve\_description\textgreater

$\{$CVE\_DESCRIPTION$\}$

\textless/cve\_description\textgreater

When processing the description, follow these steps: 

1. Thoroughly read the entire CVE description to understand the information about affected repositories and fixed version numbers.

2. Extract the names of all affected repositories and their corresponding fixed version numbers from the description.

3. Ensure that you identify all possible repositories and version numbers mentioned in the description.

4. Compile the extracted information into a list, with each item in the format of (repository, version number).

Please output your result in the <answer> tag.

\textless answer\textgreater

[Here you should list the (repository, version number) pairs]

\textless/answer\textgreater

\end{promptbox}
\caption{Prompt for Extracting Repository and Versions}
\Description{Prompt for Extracting Repository and Versions}
\label{fig:prompt_ver}
\end{figure}

\subsubsection{Multi-branch Cross-filtering}
In repositories maintained with multiple branches, a vulnerability patch may be committed across several branches. To reduce the interference of unrelated commits and improve detection quality, we further screen potential patch commits by analyzing the frequency of commits appearing in multiple versions. Specifically, based on our empirical findings, patch commits with the same fixed function typically appear in multiple versions, and their commit messages are generally similar. Therefore, we calculate the frequency of commit messages to identify high-frequency commits, which are generally likely to be patches that fix vulnerabilities.

Combining the sections above, the specific algorithmic steps in the Filtering Candidate Commits phase are as follows. The first two steps involve version acquisition, while the latter two steps apply Multi-branch Cross-filtering:
\begin{enumerate}
    \item Match the fixed version number from the vulnerability information with the repository version tags.
    \item Retrieve all commits between the previous version and the fixed version.
    \item Calculate the frequency of each commit's occurrence across all affected versions.
    \item Select the commit with the highest frequency as the candidate patch commit.
\end{enumerate}

\subsection{Multi-turn Dialogue and Voting}
In this phase, relevant function information related to patch modifications is derived using code analysis tools. Subsequently, prompts are constructed from CVE information, commit information, and function information to input into a LLM, utilizing a multi-turn dialogue prompt method to manage situations where tokens are excessively long. This entire process is executed multiple times, ultimately selecting the most relevant patch commit through voting.

\subsubsection{Extraction of Critical Code Information}
Functions are the primary components of code files. Therefore, the framework first identifies and extracts function definitions, function calls, as well as their corresponding parameters and return value types within the code. By extracting and analyzing this information, the framework effectively captures key information in code segments related to vulnerability repair. 

Using Antlr4 \cite{antlr4_github}, we have implemented a method to locate the corresponding function body based on modified lines. The main process involves parsing and constructing an abstract syntax tree (AST), locating the node where the modification line resides, and then traversing upwards to find the function declaration node, finally yielding the function declaration and corresponding function body. We have achieved parsing for six languages (C, C++, Java, Golang, Python, JavaScript). Therefore, for each submitted patch, we first determine the programming language of the modification based on the file name. After identifying the language, we employ tools to analyze the code and retrieve the function declarations and bodies associated with the patch modifications.

\subsubsection{Prompt Construction}
At this step, the method leverages a COT \cite{wei2022chain} prompt template to integrate key information extracted from code commit segments with existing vulnerability description information and commit details. By constructing thought chain prompts, the LLM is guided in recognizing patches. Moreover, a few-shot approach \cite{brown2020language, parnami2022learning} is adopted, incorporating an answer sample to lead the model to produce patch commit information in a predetermined format that meets the required standards, thereby enhancing the accuracy and consistency of answers. 

The prompt primarily comprises four section: \textbf{the question}, \textbf{vulnerability information}, \textbf{test commit information}, and \textbf{guidance patch commit detection example}. The prompt words for \textbf{the question} part and \textbf{the guidance patch commit detection example} part have been pre-generated. Through multi-question comparison detection output, the most effective question part prompt is identified. The guidance patch commit detection example part standardizes the output format of the conversational LLM, ensuring that the model's answers follow the reasoning process of the chain of thought, thus more accurately identifying test code commits. \textbf{The vulnerability information} section includes the CVE-ID of the vulnerability to be detected and its detailed description; \textbf{test commits} are obtained through the version commit filtering and frequency screening method proposed in this article. \textbf{The commit information} contains basic commit content, supplementary commit information, and key information within the code snippets.

% \begin{figure}
%     \centering
%     \includegraphics[width=1.0\linewidth]{pic/prompt example.pdf}
%     \caption{Vulnerability CVE-2023-46223 Prompt Example}
%     \label{fig:prompt_example}
% \end{figure}

\subsubsection{Multi-turn Dialogue and Majority Voting}
If the prompt exceeds the token limit of the model, we consider using a batched multi-turn dialogue approach to address this issue. First, a default upper limit for the number of batch commits (such as 10 entries) is set. For each batch, the information is integrated and a complete prompt is constructed in the same manner, submitted to the LLM, and answers are acquired. Patch commit information is then extracted from each batch's answers, all batch results are compiled, and LLM inquiries continue until the final patch commit is confirmed. 

However, LLMs may still exhibit unstable answers. To mitigate this, we adopt the concept of majority voting, collecting multiple responses (10+ rounds) to alleviate inconsistencies in the LLM. Specifically, we extract commit numbers from each model response and calculate the frequency of each commit, finally selecting the most frequently recognized commit as the patch. The entire process is shown in the pseudo-Python code in Algorithm \ref{algo:algorithm}.

\begin{algorithm}[t]
\LinesNumbered
\KwIn{
\begin{tabular}[t]{@{}l@{}}
$prompt$: Constructed prompt \\
$candidate$: A set of candidate commits \\
\end{tabular}
}
\KwOut{$patch\_commits$: A set of patch commits}
batch\_size = 10 \\
rounds = 10 \\

selected\_commits = [] \\
\For{round \textbf{in} \textbf{range}(rounds)}{
    \lightcomment{\# Multi-turn Dialogue} \\
    commits = $candidate$.copy() \\
    \While{\textbf{len}(commits) > batch\_size}{
        processed\_commits = [] \\
        \For{i \textbf{in} \textbf{range}(0, len(commits), batch\_size)}{
            batch = commits[i : i+batch\_size] \\
            commit = askLLM($prompt$, batch) \\
            processed\_commits.append(commit) \\
        }
        commits = processed\_commits \\
    }
    selected\_commits.extend(commits) \\
}
\lightcomment{\# Majority Voting} \\ 
counter = Counter(selected\_commits) \\
max\_count = \textbf{max}(counter.values()) \\
$patch\_commits$ = [item \textbf{for} item, count \textbf{in} counter.items() \textbf{if} count == max\_count] \\

\Return{$patch\_commites$}
   
\caption{Multi-turn Dialogue and Majority Voting}
\label{algo:algorithm}
\end{algorithm}

% \begin{algorithm}[h]
% \caption{algorithm caption}%算法名字
% \LinesNumbered %要求显示行号
% \KwIn{input parameters A, B, C}%输入参数
% \KwOut{output result}%输出
% some description\; %\;用于换行
% \For{condition}{
% 　　only if
% 　　\If{condition}{
% 　　　　1
% 　　}
% }
% \end{algorithm}

Finally, to improve recall, we search for code commits related to the resultant code commit, expanding the final patch commit result set. Specifically, code commits from two versions close to the repair version are used as a retrieval enhancement database. This article considers a commit as related under any of the following conditions: 
\begin{enumerate}
    \item The resultant code commit and the retrieval commit have consistent code modification segments.
    \item Retrieval commit titles and contents share identical, containing, or being contained relationships with the resultant code commit titles and contents.
    \item CVE descriptors or ISSUEID corresponding to the CVE vulnerability exist within the commit information.
\end{enumerate}

This multi-faceted approach addresses the inherent variability in language model outputs while maximizing the detection of relevant patch commits. By implementing majority voting across multiple model responses and expanding the result set through relationship-based retrieval, the framework achieves both high precision and recall in identifying vulnerability patches even in complex software repositories with numerous commits and interdependent code modifications.

\section{Evaluation}
\label{sec:eval}
To evaluate the effectiveness of our proposed two-stage vulnerability patch detection framework, this chapter conducts a series of experiments to validate the method's performance, efficiency improvements, and practical application value. Our evaluation focuses on performance comparison with existing optimal methods, contribution analysis of framework components, and practical usability verification. We aim to answer the following research questions:
\begin{itemize}
    \item \textbf{RQ1:} Does our proposed two-stage framework demonstrate significant improvements in vulnerability patch identification accuracy compared to existing optimal methods?
    \item \textbf{RQ2:} Can our method maintain effective identification performance when processing entirely new vulnerabilities that emerged after the training data cutoff time?
\end{itemize}

\subsection{Experimental Setup}
To compare with previous methods, we conducted experimental validation using a real test dataset containing 750 open-source software vulnerabilities, which serves as a common test set for all methods. We selected Tracer as the primary comparison baseline because, according to the analysis in Section \ref{sec:backgroud}, Tracer performs optimally among existing methods and represents the highest level of knowledge-driven approaches.

In our method, the main hyperparameters are the number of queries and the batch size per query, corresponding to "batch\_size" and "rounds" in lines 1 and 2 of Algorithm \ref{algo:algorithm}. In the experiments, these two values are set to 10 respectively, and we use GPT-4o-mini as the LLM with training data cutoff at October 2023. For baselines, we use the default hyperparameter settings from the open-source code. Experiments are conducted on a machine equipped with an i9-12900K CPU, two Nvidia GeForce RTX 3090 GPUs, and 96GB of memory.

\begin{table}[htbp]
\centering
\caption{Method Performance Comparison Results}
\label{tab:method_comparison}
\begin{tabular}{|l|c|c|c|c|}
\hline
\textbf{Method} & \textbf{Precision} & \textbf{Recall} & \textbf{F1} & \textbf{Accuracy (\%)} \\
\hline
Tracer & 0.5303 & 0.5734 & 0.5235 & 472 (62.9\%) \\
\hline
Ours & 0.7720 & 0.6475 & 0.6598 & 627 (83.6\%) \\
\hline
\end{tabular}
\end{table}

\subsection{Overall Performance Evaluation}
\textbf{RQ1:} Does our proposed two-stage framework demonstrate significant improvements in vulnerability patch identification accuracy compared to existing optimal methods?

Table \ref{tab:method_comparison} presents the detailed performance comparison results between our method and Tracer. From the experimental results, we observe that our method achieves significant improvements across all key metrics. Specifically, precision improves from 53.03\% to 77.20\%, representing a 45.6\% improvement; recall increases from 57.34\% to 64.75\%, showing a 12.9\% improvement; and F1-score rises from 52.35\% to 65.98\%, indicating a 26.0\% overall performance enhancement.
More importantly, our method can correctly predict 627 CVEs (83.6\%) compared to Tracer's 472 (62.9\%), representing a 32.9\% improvement. This result demonstrates that in practical application scenarios, system administrators would only need to manually verify 16.4\% of cases, significantly reducing manual workload and providing substantial practical value.
The significant precision improvement primarily benefits from our version filtering strategy, which effectively avoids false positives among numerous irrelevant commits by reducing the search space. The stable recall improvement stems from the powerful semantic understanding capabilities of large language models, which can identify complex patch patterns that traditional methods might overlook.

\begin{table}[htbp]
\centering
\caption{CVE-2024 Method Performance Comparison Results}
\label{tab:cve-2024_performace}
\begin{tabular}{|l|c|c|c|c|}
\hline
\textbf{Method} & \textbf{Precision} & \textbf{Recall} & \textbf{F1} & \textbf{Accuracy (\%)} \\
\hline
Tracer & 0.5233 & 0.6400 & 0.5560 & 16 (64.0\%) \\
\hline
Ours & 0.7333 & 0.8533 & 0.7600 & 22 (88.0\%) \\
\hline
\end{tabular}
\end{table}

\subsection{Timeliness Analysis}
\textbf{RQ2:} Can our method maintain effective identification performance when processing entirely new vulnerabilities that emerged after the training data cutoff time?

To validate the timeliness and generalization capability of our method for unknown vulnerabilities, we designed a rigorous temporal split experiment. Considering that our large language model GPT-4o-mini has a training data cutoff of October 2023, we selected 25 CVEs with CVE-2024 prefixes from our test set for specialized analysis. The official disclosure dates of these vulnerabilities are all after January 1, 2024, ensuring that related vulnerability information is completely absent from the model's training data. Additionally, by querying the release times of patch commits corresponding to these vulnerabilities, we confirmed that all patch commit times are also later than the model's training data cutoff, thus constituting a genuine "future data" test set.

Table \ref{tab:cve-2024_performace} presents the performance comparison results on these 25 entirely new vulnerabilities. Remarkably, our method still demonstrates excellent performance when handling these completely unknown vulnerabilities. This result has important practical implications. First, it proves that our method possesses strong generalization capability, maintaining high-level identification performance even when facing entirely new vulnerability types and patch patterns that the model has never "seen." This generalization capability primarily stems from our method's design philosophy: the strategy of using version information filtering to reduce search space has universal applicability and does not depend on specific vulnerability knowledge; while the large language model, although it has not encountered specific vulnerability instances, can identify new vulnerability-patch mapping patterns through the general code understanding and reasoning capabilities learned from large-scale code corpora.

\section{Discussion}
\label{sec:discussion}
%% 版本搜索的限制
%% 数据获取的限制
%% commit的文件过滤
This section provides an in-depth analysis of the core mechanisms, technical contributions, and practical application considerations of our method, aiming to offer deeper understanding and guidance for future development in the vulnerability patch detection field.

\textbf{Deep Mechanism Analysis of Method Effectiveness}.
The successful application of large language models demonstrates the advantages of knowledge-driven methods in specialized tasks. Unlike traditional methods that attempt to learn complex mapping relationships from limited samples, LLMs have already internalized rich code understanding and reasoning capabilities through pre-training. This capability enables them to handle complex situations that traditional methods struggle with, such as subtle logic repairs and cross-file associated modifications.

\textbf{Important Significance of Generalization Capability and Timeliness}.
Our method demonstrates certain generalization capabilities, which primarily stem from the universal design of the approach: the version filtering strategy is based on universal principles of software engineering and does not depend on specific vulnerability knowledge; the semantic understanding capability of LLMs enables them to identify new vulnerability-patch patterns. This design philosophy provides reliable safeguards for handling rapidly evolving security threats. Moreover, when testing "entirely new vulnerabilities" for the LLM, our method still maintains high performance.

\textbf{Considerations and Limitations for Practical Deployment}.
Despite significant improvements achieved by our method, several important factors must be considered in practical deployment. While large language models provide powerful analytical capabilities, their computational costs are relatively high. In large-scale deployments, a balance must be found between accuracy and cost. Although our two-stage design reduces the number of candidates that need LLM processing, somewhat alleviating this issue, the version information identified based on CVE descriptions may not always pinpoint patch locations with 100\% accuracy, leaving room for further reduction.

\textbf{Future Research Directions}.
Based on the findings and limitations of this study, we believe future research should focus on \textbf{web information fusion}. While we primarily focus on code and version information, enabling large models to act as agents that automatically search the web for relevant information to assist vulnerability patch localization should allow them to discover vulnerability patch information from subtle clues, similar to how engineers conduct their analysis.

\section{Related Works}
\label{sec:related}
There are numerous studies focusing on the localization of security patches, which can be categorized into two types: \textbf{tracing security patches for disclosed vulnerabilities} and \textbf{identifying silent security patches}.

For \textbf{tracking security patches of disclosed vulnerabilities} \cite{nguyen_vulcurator_2022, nguyen-truong_hermes_2022, tan_locating_2021, wang_vcmatch_2022, xu_tracking_2022, li_patchfinder_2024, zhang_dual_2024}. Xu et al. \cite{xu_tracking_2022} conducted an empirical study to understand the quality and characteristics of patches for disclosed vulnerabilities in two industrial vulnerability databases, thus proposing to track patches from CVE reference links across multiple knowledge sources (e.g., Debian). Their research concentrates on analyzing reference links provided by security analysts, instead of directly tracing patches from OSS repositories. PATCHSCOUT \cite{tan_locating_2021} attempted to leverage the correlations between vulnerability descriptions and code commits to locate patches, using RankNet \cite{burges2010ranknet} based on handcrafted features between vulnerability descriptions and code commits to identify security patches. As PLMs in NLP have evolved, researchers have attempted to use PLMs to identify security patches and establish the correlation between vulnerability descriptions and commits. VCMatch \cite{wang_vcmatch_2022} directly classifies a commit as related or unrelated to the CVE description by fusing features from PatchScout and vectors extracted from Bert. Nguyen et al. \cite{nguyen_vulcurator_2022} introduced a security patch location tool that uses a CodeBERT \cite{feng2020codebert} to encode the commit message, code changes, and issue reports, generating the probability that the commit is a security patch. PatchFinder \cite{li_patchfinder_2024} introduced a two-phase framework designed to overcome challenges posed by large search spaces and enables end-to-end fine-tuning to fully exploit the natural correlation between CVE descriptions and commits. PromVPat \cite{zhang_dual_2024} uses dual prompt tuning channels to capture semantic correlations between vulnerability descriptions and code commits, enhancing the method's performance in few-shot scenarios. Regarding LLMs, Fu et al. \cite{fu_chatgpt_2023} evaluating the performance of LLMs (ChatGPT) in tasks related to software vulnerability detection. However, these works primarily focus on predicting and classifying vulnerabilities, rather than locating patches. In our work, we directly utilize multi-round dialogue prompts from LLMs, combined with version localization to filter out unrelated code commits, to locate the security patches for specific vulnerabilities defined by CVE descriptions without the need for model fine-tuning.

In studies focusing on \textbf{identifying silent security patches} \cite{wang_detecting_2019, xu_spain_2017, cabrera_lozoya_commit2vec_2021, wang_patchrnn_2021, wu_enhancing_2022, zhou_colefunda_2023, zhou_finding_2021, zhou_automated_2017, zhou_spi_2021}, the aim is to find parts of code commits that constitute security patches without associating them with specific vulnerabilities they rectify. Early studies identified updated patches based on summarizing patterns of patches and vulnerabilities. For example, SPAIN \cite{xu_spain_2017} identified and searched for similar patches or vulnerabilities by comparing the binary programs of the original and patched versions, using summarized patterns. Later research began to use machine learning techniques to identify patches. For example, Wang et al. \cite{wang_detecting_2019} utilized machine learning by building features to automatically identify silent security patches in open-source software. PatchRNN \cite{wang_patchrnn_2021} uses recurrent neural networks to identify security patches in OSS updates. Zhou et al. \cite{zhou_finding_2021} employed CodeBERT \cite{feng2020codebert} to extract semantic representations from code changes and identify whether the input code changes are silent security patches. CoLeFunDa \cite{zhou_colefunda_2023}, based on function change information, first enhances data and then uses contrastive learning to learn change representations, finally achieving the identification of patches and judgment of vulnerabilities the patch rectifies. In contrast, our focus is on tracing security patches tailored to a particular vulnerability, as defined by its CVE description.

\section{Conclusion}
\label{sec:conclusion}
This study reveals four key limitations of existing vulnerability patch detection methods through systematic analysis: oversized search spaces, complex architectures underperforming pre-trained semantic understanding, timeliness constraints of crawler-based approaches, and the superiority of knowledge-driven over feature-based methods.

Based on these insights, we propose a two-stage framework that combines version-driven filtering with LLM-based dialogue voting to achieve accurate, efficient, and real-time vulnerability patch detection. Experimental results demonstrate significant performance improvements and strong generalization capabilities on entirely new vulnerabilities, validating the effectiveness of our knowledge-driven approach. This work establishes a new paradigm for vulnerability patch detection that prioritizes semantic understanding over architectural complexity, providing a foundation for more intelligent and responsive security maintenance tools.

\section{Data Availability}
Data were collected from CVE \cite{cve}, NVD \cite{nvd2025}, Vera \cite{vera}, and Snyk \cite{snyk_vuln_db} databases. We reproduced four baseline methods: VCmatch \cite{vcmatch_code}, PromVPat \cite{dual_package_2023}, PatchFinder \cite{patchfinder_code}, and Tracer \cite{tracer_code_2021}. The complete experimental materials, including processed datasets, our proposed method implementation, and reproduced baseline code, are made publicly available at https://anonymous.4open.science \cite{VulnerCollector_2023} to ensure reproducibility.

\bibliographystyle{plain}
\balance
\bibliography{main}

\end{document}